\documentclass[pra,superscriptaddress,twocolumn,superscriptaddress,amsmath]{revtex4}

\usepackage{amsmath, amsthm, amssymb}
\usepackage[]{graphicx} 
\usepackage{dcolumn} 
\usepackage{bm} 
\usepackage[normalem]{ulem}
\usepackage{cases}
\usepackage{color}
\usepackage[dvipsnames]{xcolor}
\usepackage{algorithm}
\usepackage{algpseudocode}
\usepackage{csquotes}
\usepackage{mathtools}
\usepackage{mathrsfs}  
\usepackage{mathtools}
\usepackage[caption=false]{subfig}
\usepackage{lipsum}
\usepackage{comment}
\usepackage{tabularx}
\usepackage{multirow}
\usepackage{braket}
\usepackage{pgf} 
\usepackage{siunitx} 

\usepackage{hyperref}
\hypersetup{
    colorlinks=true,
    linkcolor=blue,
    citecolor=blue,
    filecolor=magenta,      
    urlcolor=cyan
    }

\begin{document}

\title{Intermodal quantum key distribution field trial \\ with active switching between fiber and free-space channels}

\author{Francesco~Picciariello}
\author{Ilektra~Karakosta-Amarantidou}
\author{Edoardo~Rossi}
\author{Marco~Avesani}
\author{Giulio~Foletto}
\affiliation{Dipartimento di Ingegneria dell'Informazione, Universit\`a degli Studi di Padova, via Gradenigo 6B, IT-35131 Padova, Italy}

\author{Luca~Calderaro}
\affiliation{ThinkQuantum s.r.l., via della Tecnica 85, IT-36030 Sarcedo, Italy}

\author{Giuseppe~Vallone}
\affiliation{Dipartimento di Ingegneria dell'Informazione, Universit\`a degli Studi di Padova, via Gradenigo 6B, IT-35131 Padova, Italy}
\affiliation{Padua Quantum Technologies Research Center, Universit\`a degli Studi di Padova, via Gradenigo 6A, IT-35131 Padova, Italy}

\author{Paolo~Villoresi}
\affiliation{Dipartimento di Ingegneria dell'Informazione, Universit\`a degli Studi di Padova, via Gradenigo 6B, IT-35131 Padova, Italy}
\affiliation{Padua Quantum Technologies Research Center, Universit\`a degli Studi di Padova, via Gradenigo 6A, IT-35131 Padova, Italy}

\author{Francesco~Vedovato}
\email[Corresponding author: ]{francesco.vedovato@unipd.it}
\affiliation{Dipartimento di Ingegneria dell'Informazione, Universit\`a degli Studi di Padova, via Gradenigo 6B, IT-35131 Padova, Italy}
\affiliation{Padua Quantum Technologies Research Center, Universit\`a degli Studi di Padova, via Gradenigo 6A, IT-35131 Padova, Italy}

\date{\today}

\begin{abstract}
{\textbf{Background:} Intermodal quantum key distribution  enables the full interoperability of fiber networks and free-space channels, which are both necessary elements for the development of a global quantum network. We present  a field trial of an intermodal quantum key distribution system in a simple 3-node heterogeneous quantum network --- comprised of two polarization-based transmitters and a single receiver --- in which the active channel is alternately switched between a free-space link of 620~m and a 17km-long deployed fiber in the metropolitan area of Padova. \textbf{Findings:}  The performance of the free-space channel is evaluated against the  atmospheric turbulence strength of the link. The field trial lasted for several hours in daylight conditions, attesting the interoperability between fiber and free-space channels, with a secret key rate of the order of kbps for both the channels. \textbf{Conclusions:} The quantum key distribution hardware and software require no different strategies to work over the two channels, even if the intrinsic characteristics of the links are clearly different. The switching system represents a cost-effective solution for a trusted quantum key distribution network, reducing the number of necessary devices in different network topologies.}
\end{abstract}

\maketitle

\section{Introduction}

Quantum key distribution (QKD) has rapidly developed toward commercial systems to meet the requirements and interests of different final users for a global quantum infrastructure, namely a quantum network. In the case of small urban-scale~\cite{Peev_2009, sasaki:11,Chen:09} or inter-urban quantum networks, the connection between nodes is usually realized with telecommunication fibers. While fiber-based QKD is the optimal choice for short distances between nodes, the presence of optical losses over long fiber links or the unavailability of an accessible deployed fiber requires the use of trusted relays~\cite{chen2021integrated} or of an alternative QKD medium, such as a free-space channel between optical telescopes.

 A fiber link might not always be viable for physical, political, or economic constraints, thus rendering free-space QKD necessary. Additionally, fiber-based QKD does not encompass all the possible terrestrial QKD needs. For example, free-space QKD can be exploited for rural areas connections, ship-to-ship or ship-to-port links, cross-border links, or to deploy a portable or relocatable network for events like summits and conferences. Free-space QKD has already been demonstrated for satellite-to-ground~\cite{liao2017satellite, Li:22}, ground-to-ground links including urban and daylight scenarios where the atmospheric turbulence is generally more severe~\cite{peloso2009daylight, gong2018free, avesani2021fulldaylight, basset2023daylight, gruneisen2021adaptive, krvzivc2023towards}, and with moving platforms as  drones~\cite{tian2023dronebased}, airplanes~\cite{airplane} and balloons~\cite{balloon}. However, typically free-space QKD requires the adoption of ad-hoc QKD hardware complementing the telescopes, thus being incompatible with commercially available QKD systems. Hence, the true interoperability between fiber and free-space links, that is necessary for the development of a fully integrated quantum network, has yet to be achieved.

 A possible solution to make free-space channels compatible with commercial fiber-based QKD devices is to exploit a hybrid free-space/fiber intermodal channel. Today, the majority of commercial QKD products share some features, as a signal wavelength around 1550~nm, the use of one single-mode-fiber (SMF) to implement the quantum channel, the need of another service channel (e.g., for the authenticated classical channel or synchronization purposes) realized via ethernet or via an additional SMF, and the operation up to a certain amount of attenuation. The core idea of intermodal QKD (IM-QKD) is to take advantage of these similarities by using directly the QKD devices for the free-space link without any modifications, connecting the QKD transmitter (Alice) to the transmitting telescope and the QKD receiver (Bob) to the receiving one~\cite{vedovato2023realization}. While the former can be straightforward, the latter needs to couple the optical free-space signal into a SMF at the receiving telescope, a challenging task that can be achieved by exploiting Adaptive Optics (AO) technology~\cite{tyson2022principles}. 
 As a result, and given that an appropriate AO system is implemented, an intermodal channel can be used whenever a fiber is not viable, and it may be useful when different QKD transmitters or receivers share the same optical telescopes or when it is difficult to establish a trusted security perimeter around them, since the QKD devices may be located elsewhere with respect to the telescopes. The feasibility of IM-QKD for point-to-point communication (i.e., one Alice and one Bob) along a hybrid free-space/fiber channel has been demonstrated in two field trials in the cities of Vienna and Padova~\cite{vedovato2023realization}.

\begin{figure*}
    \centering
    \includegraphics[width=\textwidth]{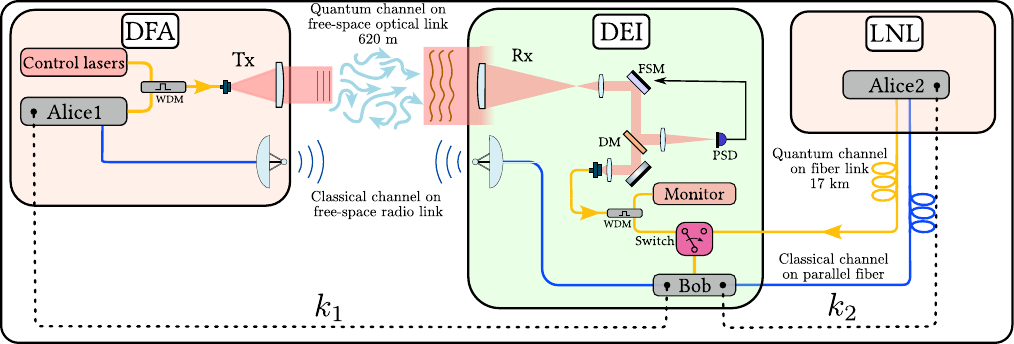}
    \caption{System architecture of the field trial. The key $k_1$ shared by Alice1 and Bob is generated exploiting a free-space quantum channel of around 620~m between DFA and DEI. The key $k_2$ shared by Alice2 and Bob is generated exploiting a deployed-fiber quantum channel that connects DEI and LNL. Before the quantum receiver, Bob, an optical switch selects the  active link. The classical post-processing runs on free-space radio link between Alice1 and Bob and on an optical fiber parallel to the one used for the quantum communication between Alice2 and Bob.WDM: wavelength-division-multiplexer.}
    \label{fig:system_arch}
\end{figure*}

In this work, we extend the use of the IM-QKD method by implementing for the first time a simple 3-node heterogeneous free-space/fiber network comprised of two QKD transmitters (Alice1 and Alice2) and a single QKD receiver (Bob). Our implementation is based on an optical switch (as used in previous homogeneous multi-node fiber networks~\cite{chen2010metropolitan,chang2016experimental,toliver2003experimental,elliott2005current}) alternating between a free-space channel of 620~m (Alice1-Bob) and a deployed fiber channel of 17~km (Alice2-Bob, see Fig.~\ref{fig:system_arch}), demonstrating the full interoperability of the two different channels. Regarding the used QKD hardware, we exploited both QKD systems realized in-house and commercially available solutions using polarization encoding at 1550~nm. 

The novelty of this work lies in the fact that the implemented links are heterogeneous (fiber and free-space), while the QKD terminals were designed for fiber links. In particular, the two different channels share the same QKD receiver (Bob). We stress that, from the point of view of the QKD hardware and processing, there are no actual differences between the two channels and that the same IM-QKD technique can be adapted to a greater number of nodes, as expected in a fully integrated complex free-space/fiber network. We report secret key rates (SKRs) of the order of kbps under daylight conditions, achieved with the aid of free-space optical terminals that are versatile, robust, compact, and easy to integrate with the fiber infrastructure. Moreover through these trials, we validated the model proposed to optimally design free-space quantum key distribution systems~\cite{scriminich2022optimal}. 

\section{Methods}

\subsection{System Architecture, QKD devices and switch}

The IM-QKD field trial that is presented in this work consisted of three nodes of the Padova University Network: one at the Department of Physics and Astronomy (DFA), one at the Department of Information Engineering (DEI) and one at the National Institute for Nuclear Physics (INFN) in Legnaro (LNL).

The system architecture of the trial is presented in Fig.~\ref{fig:system_arch}.  Two QKD transmitters were placed in DFA and LNL (Alice1 and Alice2) and a QKD receiver in the DEI node (Bob). For the free-space link between DFA and DEI, Alice1 and Bob exploit the SMF-injection realized at the receiving telescope placed in the same building of Bob, while for the fiber link between LNL and DEI Alice2 and Bob are directly connected through a deployed cable. 

Regarding the QKD devices, in our implementation Alice1 and Bob were commercial QKD devices (QUKY, ThinkQuantum srl) while Alice2 was realized in-house at the University of Padova. Both systems exploit the efficient BB84 protocol
with 3-states and 1-decoy technique~\cite{Rusca2018} with polarization encoding for the generation of the qubits and the qubit-based  algorithm Qubit4Sync of Ref.~\cite{Calderaro2020} to perform the synchronization between the QKD transmitter and receiver. Regarding the optical schemes of the QKD transmitters, a complete description of Alice2 can be found in Ref.~\cite{Avesani2021resource}. In summary, for Alice2, the quantum optical signal at 1550.12~nm (ITU-grid channel 34) is generated with a gain-switched distributed feedback (DFB) laser, with a repetition rate of 50~MHz and a 270~ps pulse duration. The qubits are encoded in the polarization of the photons using an iPOGNAC polarization encoder based on a Sagnac interferometer, described in details in Refs.~\cite{Agnesi:19, Avesani:iPOGNAC} and to implement the decoy states method, a beam-splitter-based intensity modulator has been realized following Ref.~\cite{Roberts2018}. The optical scheme of Alice1 is similar to the one of Alice2, since Alice1 also presents a DFB laser running at 50~MHz of repetition rate at the wavelength of 1550.12~nm, followed by an intensity mo\-dulation stage to apply the two different intensities required by the 1-decoy technique and an iPOGNAC encoder for the polarization modulation of the emitted qubits. The quantum receiver, Bob, implements a passive decoder with just one single-photon-detector, with an optical scheme similar to the one of Ref.~\cite{avesani2022deployment}. 

A switching scenario between the two transmitters was implemented; in particular, using an optical switch before Bob at the DEI node, we tested the free-space and fiber channels in a cycling manner of 15 minute-long intervals. At each switch, Bob uses automatic polarization controllers to align its measurement basis to the one used by Alice, which sends fixed and known states for a brief preliminary phase before beginning the key exchange.
When the quantum bit error rate (QBER) is sufficiently low, the key exchange begins, but alignment also continues using publicly known sequences of bits that are interleaved with the actual key~\cite{avesani2022deployment}.

We employed a custom-developed network-controlled optical switch. The device is composed by a 2$\times$2 fiber switch with approximately
$-1$~dB of insertion loss and it is connected to a microcontroller that handles its operation and exposes a serial interface. When remote operation is required, the microcontroller is paired with a System on Module (SoM), which allows one to operate the device over the network via an IP-based socket system. The device's software has been developed and integrated into the QKD's control software allowing both manual and automatic operation. As an example, the switching between the two channels can be either triggered by the user or by the control software when certain conditions (such as elapsed time) are met. 
In this sense, the choice for a 15-minute switch cycle was made for demonstrative purposes, and it can and should be adjusted depending on the actual network implementation. For example, in the case of unstable free-space channel conditions, the switching can be associated to a degradation of the channel, to avoid dead time in the key generation.

From a quantum network perspective, we equipped our system with a quantum logic layer consisting in a key manager system compliant with ETSI standards for the communication with the application layer, similarly to the experiment described in Ref.~\cite{picciariello2024quantum}.

\subsection{Description of the intermodal system}

The intermodal system was based on two main telescopes and a pointing, acquisition and tracking (PAT) system, to guarantee the fine alignment and the SMF coupling of the signal at the receiving telescope. 

The free-space transmitter (Tx) used in the trial is comprised of a terminated SMF fiber channel/physical contact (FC/PC) adapter mounted on a linear stage and placed in the back focus of a 2-inch lens to produce, after its collimation, a beam with a waist of about 25~mm. Beyond the QKD signal at 1550~nm provided by Alice1, two additional beacon lasers, one at 980~nm and one at 1545~nm, are combined into the same SMF via wavelength multiplexing. These two beacons are exploited by the fine-alignment process between the free-space Tx and Rx terminals, implemented at the receiver side.

The SMF at the receiver (Rx) acts as a spatial filter to achieve daylight operation, and the coupling of the signals in the fiber allows the integration of QKD systems with the standard telecommunication infrastructure. However, a reliable fiber coupling imposes restrictions on the design of the free-space terminal. This is true since there are detrimental impairments caused by the propagation of an optical beam through atmospheric turbulence, namely scintillation, beam broadening, wandering and wavefront distortion, that turn SMF coupling into a challenging task. To mitigate the latter, one option is to employ adaptive optics (AO) techniques at the receiver side~\cite{gruneisen2021adaptive}. We designed the free-space system by following the procedure described by Scriminich et al. in Ref.~\cite{scriminich2022optimal}, with a target working distance of about 1~km and by making the terminals as compact and light as possible for portability and ease of installation. We report here some useful considerations on the terminal design.

In the case of a horizontal free-space link with fiber-coupling at the receiver, the overall channel efficiency $\eta_{\rm CH}$ is given by
$\eta_{\rm CH}=\ \eta_{\rm A}\eta_{D_{\rm Rx}}\eta_{\rm SMF}$~\cite{scriminich2022optimal}, 
where $\eta_{\rm A}$ denotes the atmospheric channel absorption, $\eta_{D_{\rm Rx}}$ the receiver collection efficiency and $\eta_{\rm SMF}$ the SMF coupling efficiency. Since the absorption coefficient at the working wavelength $\lambda =$1550 nm is less than $-1$~dB/km~\cite{scriminich2022optimal}, the introduced attenuation is negligible with respect to the other terms and is not considered. 

The collection efficiency $\eta_{D_{\rm Rx}}$ depends on the ratio of the receiver aperture $D_{\rm Rx}$ to the beam-waist radius at the receiver. To properly address the influence of the atmosphere and thus designing the optical receiver we referred to the estimation of the refractive index structure constant $C_n^2 \simeq 2 \cdot 10^{-13}$~m$^{-2/3}$ reported for a previous experiment realized in Padova~\cite{avesani2021fulldaylight}. For a transmitter producing a Gaussian beam waist of about 25~mm, the beam-waist radius broadened by turbulence after a 1 km long propagation becomes approximately 90~mm. Hence, the expected collection losses for a receiver with $D_{\rm Rx} = 50.8$~mm place at a 1~km distance from the transmitter is around~$-8$~dB. 

The average SMF coupling efficiency $\eta_{\rm SMF} = \eta_0\eta_{\rm AO}\eta_{\rm S}$ of the beam collected by the telescope aperture is a product of the optical efficiency $\eta_0$ of the Rx telescope, the efficiency of the AO system $\eta_{\rm AO}$, and a term $\eta_S$ due to atmospheric scintillation. 
The term due to scintillation is at most $-1$~dB for all turbulence regimes~\cite{scriminich2022optimal}. The optical efficiency $\eta_0$ can be obtained by knowing the obstruction ratio of the receiver aperture and in the optimal case of no obstruction one can achieve an efficiency of at most $\eta_0\approx\ 81.5\%\ \approx\ -0.89$ dB.

The efficiency $\eta_{\rm AO}$ depends on the performance of the AO system implemented and on the expected strength of the turbulence present in the channel. Typically, the turbulence strength is parameterized by the ratio $D_{\rm Rx}/r_0$, with $r_0$ the Fried parameter~\cite{noll1976zernike, scriminich2022optimal, fried1965statistics}, since the variances of the Zernike coefficients, which are used to decompose the wavefront into different aberrations, are analogous to this term. On the other hand, the effectiveness of the AO system is captured by the maximum obtainable order of correction~\cite{noll1976zernike}.

From the estimated value of $C_n^2$ and the wavevector $k=2\pi/\lambda$, we can compute the Fried parameter for a link of length $L=1$~km as $r_0 = \left( 0.42 \ C_n^2 k^2 L \right)^{-3/5} \simeq 13$~mm~\cite{tyson2022principles}. Hence, the expected ratio $D_{\rm Rx}/r_0$ is around 4 and an effective improvement of the SMF injection efficiency with respect to the case of no correction is obtained by performing a mere tip-tilt correction. 
Since the losses before the AO system add up to approximately 10~dB, we estimate that our system can operate effectively under turbulence conditions corresponding to $D_{\rm Rx}/r_0 \lesssim 7$ (see Fig.~6 of Ref.~\cite{scriminich2022optimal}), given that the QKD devices involved in the field trial can tolerate overall channel losses of about $-20$~dB.

With these considerations in mind, the free-space Rx includes a $6\times$ beam reducer of 2-inch (50.8~mm) aperture followed by a fast-steering-mirror (FSM, model STT-25.4 by SmarAct), and a dichroic mirror (DM) to separate the 980~nm beam from the ones in the telecom C-band (QKD signal and 1545nm-beacon). The FSM is used to correct the angle of arrival fluctuations of the incoming beam, using the feedback provided by a position-sensitive-detector (PSD) placed on the focal plane of a 300~mm lens. The C-band beams are instead coupled to an FC/PC-terminated SMF. The focal length of the lens in front of the fiber is chosen to guarantee that the effective focal length $f_{\rm eff}$ of the receiving telescope with aperture $D_{\rm Rx}$ at wavelength $\lambda$ is close to the optimal value $f_{\rm eff} = \pi D_{\rm Rx} {\rm MFD} / (4 \lambda \beta) \approx 230$~mm~\cite{scriminich2022optimal},
 where $\beta\approx1.12$, since there is no obstruction in the Rx telescope aperture and the mode field diameter (MFD) of the SMF is 10.4~$\mu$m. After the fiber coupling, a 100~GHz WDM filter is used to separate the QKD signal from the 1545-nm beacon, that is used to monitor in real-time the coupling efficiency of the channel.  

 Beyond the main optical tubes, we equipped both the Tx and the Rx with a coarse telescope of 1-inch aperture and a CMOS camera with a field-of-view (full angle) of approximately 6~mrad. These two telescopes are used to roughly align the two terminals in the preliminary phases of the experiment. Both terminals use cage assemblies for a total weight of around 5~kg and are mounted on a commercial alt-azimuth mount (AZ-GTi by SkyWatcher). Two commercially available RF antennas were installed at the Tx and Rx locations to support a bidirectional classical channel.

\section{Results}

\subsection{Results for the fiber link}
\begin{figure}[t]
    \centering
    \includegraphics[width=\columnwidth]{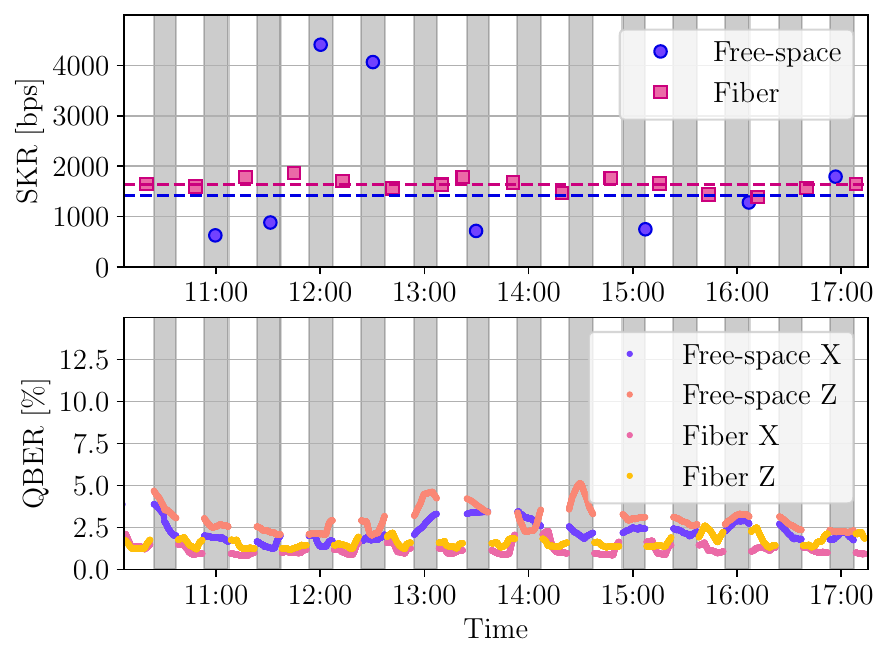}
    \caption{Achieved SKR (up) and QBER (down) during the IM-QKD field trial. The measurements concerning the fiber-based channel are marked with pink (squared markers), whilst the free-space ones with blue (circular markers). The switching interval between the two channels is approximately 15 minutes.}
    \label{fig:qber_results}
\end{figure}

The field trial took place on April 18th, 2023.
The main performance indicators for the 17km-long fiber link ($-11$~dB of attenuation) are the final SKR and QBER, reported in Fig.~\ref{fig:qber_results} (which also includes results for the free-space link).
These parameters are computed directly by the post-processing software installed on the QKD devices when it generates a secret key from raw detection data.
This software follows the specifications of the aforementioned efficient BB84 protocol with three states and one decoy level and operates on blocks of $4\cdot 10^6$ bits of sifted key. The QBER is the ratio between the number of erroneous and total detections in each basis in the sifted key.
The SKR is calculated as the ratio between the length of the secret key distilled from a block (Eq.~(1) of Ref.~\cite{Rusca2018}) and the time needed to accumulate such block.
The time needed for the post-processing itself is neglected because the procedure can be executed while the next block is accumulated.
The SKR shows remarkable stability, with values close to the average of about 1.6 kbps.
This means that the alignment algorithm has always been able to find approximately the same conditions.
This is further proven by the plot of the QBER, in which the majority of data points are well below 2\%.

It is noteworthy that at each switch from the free-space to the fiber link, the QBER is higher (close to 4\%) only to descend shortly afterward.
This is because the QBER threshold to stop the preliminary alignment procedure and start the key exchange was set to a comparatively high value, so as to keep this phase short. The live alignment correction then proceeded to further reduce the QBER, achieving an average of approximately 1.4\%.

\subsection{Results for the free-space link}
The 1545~nm beacon at the Rx was used to monitor the power coupled into the SMF concurrently with the execution of the QKD protocol in 5 minute-long acquisitions.
Through these measurements, we were able to quantify the strength of the atmospheric turbulence present in the free-space channel between Alice1 and Bob, in particular the ratio $D_{\rm Rx}/r_0$.

\begin{figure}[t!]
    \centering
    \includegraphics[width=\columnwidth]{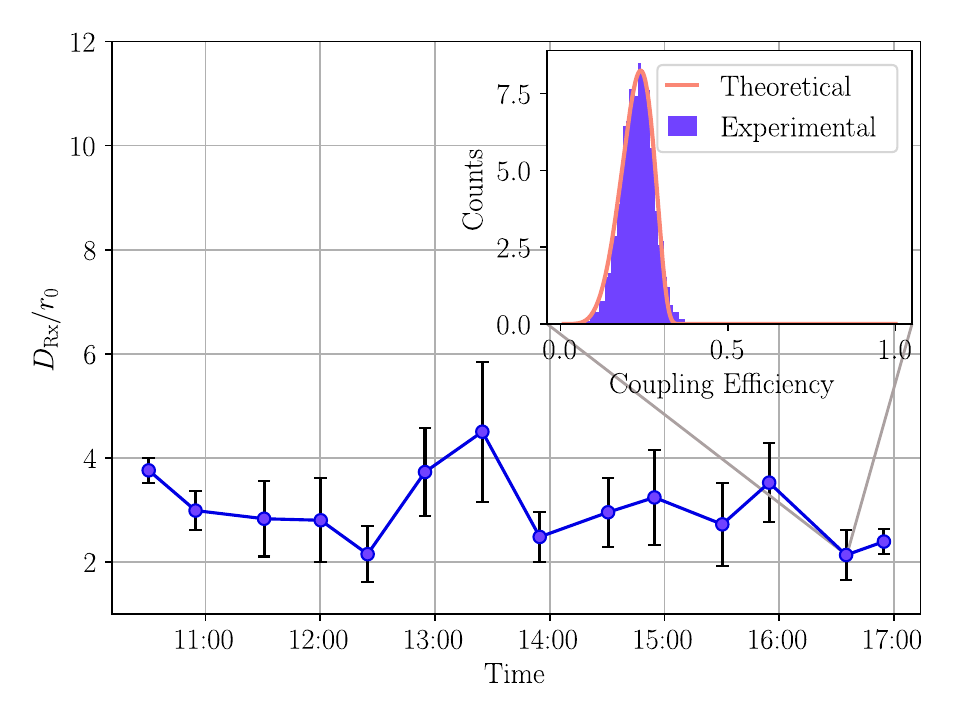}
    \caption{SMF coupling efficiency histogram when active tip/tilt correction is used. $D_{\rm Rx}/r_0$ ratio for each point extracted by fitting the theoretical SMF coupling efficiency PDF to all the power measurements acquired when the free-space channel was used for the execution of the QKD protocol (inset figure).}
    \label{fig:free_space_data}
\end{figure}

To estimate this ratio, we apply the model of the coupling efficiency probability density function (PDF) described in Ref.~\cite{canuet2018statistical}, with the modification regarding the finite-size of the control system proposed in Ref.~\cite{roddier1999adaptive}, to the measured coupling efficiency obtaining an estimation of $r_0$. 
An example of such estimation is shown in the inset of Fig.~\ref{fig:free_space_data}: This case is one of the best coupling efficiencies achieved during the field trial, corresponding to a $D_{\rm Rx}/r_0$ ratio of around 2.5. 

Applying the same analysis for all acquisitions during the key exchange, we obtain the change of $D_{\rm Rx}/r_0$ over time, shown in Fig.~\ref{fig:free_space_data}. The averaged value of the $D_{\rm Rx}/r_0$ ratio for the field trial (f.t.) is $3.0\pm0.2$, corresponding to an averaged Fried parameter $r_0^{\rm (f.t.)}$ of 17~mm and a $C_n^{2\rm(f.t.)} $ of $2.1 \cdot 10^{-13}$~m$^{-2/3}$ for the horizontal link of length $z = 620$~m. It is worth noticing that these values are aligned to the ones assumed in the design phase (see above), and they provide a Rytov variance $\sigma_R^{2{\rm (f.t.)}} = 1.23 C_n^{2\rm(f.t.)} k^{7/6} z^{11/6} = 1.7$. This estimation of the Rytov variance and the mean value of the ratio $D_{\rm Rx}/r_0$ attest that the field trial was conducted under moderate turbulence, being the weak (strong) turbulence regime conventionally described by a Rytov variance below (above) 1 and a ratio $D_{\rm Rx}/r_0$ below 3 (above 5)~\cite{Ghalaii2023,tyson2022principles}.  

We notice that almost during the entirety of the experimental run, the estimated turbulence conditions are within the range that our system can operate on ($D_{\rm Rx}/r_0 \lesssim 7$) even when taking into account the standard deviation error of the fits. The fact that we achieve positive key rates, at times higher than when the fiber-based channel is used, at the cost of a slightly increased QBER (2.5\% on average), substantiates the robustness of the design of the free-space receiver. Consequently, this experimental demonstration bolsters the model in Ref.~\cite{scriminich2022optimal} for optimally designing free-space QKD systems.

We also investigated the noise rate and found it approximately 10\% higher when using the free-space channel ($4.5 \pm 0.5$ kHz, a datum that accounts for the saturation effect due to the dead time of the detector and whose error represents the standard deviation of the noise rate sample) compared to the fiber one ($4.1 \pm 0.4$ kHz).
The majority of this noise is due to detector dark counts and afterpulses, whose amount is mostly independent of the channel choice (with afterpulses changing slightly as they reflect the total detection rate).
In the case of free-space link, the 10\% noise increase is due to solar scattered light, with detector dark counts and afterpulses remaining dominant even in full daylight, thus validating the effectiveness of the spatial and wavelength filtering operated by the receiver setup.
We calculated the direct contribution to the QBER caused by this increase to be only about 0.1\%.

As a final consideration, it is important to note that while the fiber link consistently produced at least one key block in each assigned time slot, this was not the case for the free-space link (see Fig.~\ref{fig:qber_results}). This difference is attributed to the lower raw key accumulation rate in the free-space channel, which has higher losses on average. As a consequence, more time is required to reach the necessary block size for initiating the post-processing analysis, which can result in key accumulation over multiple time slots. Generally, this does not pose a security issue. However, if there is a longer interval between consecutive slots, it is crucial to ensure that the storage solution for the raw bits is compliant with the security risk analysis of the overall system, which is a common task performed in all the classical cryptographic key management systems.

\section{Discussion}

In this work, we presented the realization of hours-long IM-QKD in a simple 3-node heterogeneous network, switching between a fiber and a free-space channel under daylight conditions. This was achieved by combining commercial and in-house developed QKD devices, and compact optical terminals with small apertures. We were able to achieve a mean SKR of around 1.5~kbps in both free-space and fiber channels, with average QBERs of 2.5\% and 1.4\% respectively.

Due to the switching functionality of the presented setup, our work can be extended to a larger number of nodes, thus constituting a building block for more complex networks. In this regard, it is worth mentioning that the most expensive components of a QKD system are the single photon detectors, and thus reducing the necessary number of Bobs (as in our implementation) represents a cost effective solution. Moreover, the free-space terminal at the transmitter side can be upgraded in order to support bidirectional optical laser communication by using a design close to the receiver's SMF injection system and wavelength-division-multiplexing technique for combining and separating the different signals, thus removing the need of external RF antennas. 
We also would highlight a possible advantage of using polarization encoding in favor of other options. For example, in the case of time-bin encoding, the calibration of the involved interferometers can be time-consuming and thus prohibitive when more than two users are involved. In our case, switching between Alice1 and Alice2 generally introduced an overhead of only 20 to 30 seconds each time. However, a thorough evaluation of the pros and cons of each encoding method should be conducted for each network configuration, especially for complex networks that may incorporate satellite links, multiplexing schemes, or more advanced QKD techniques (e.g., high-dimensional QKD~\cite{Islam2017}), for which studies and experiments are still ongoing.

Further development to be considered, starting form the presented work, are the study and validation of the analysis for longer, and typically more turbulent, free-space links. In particular, while our field trial is representative for an urban free-space link, practical applications may look forward to ranges greater than 10 km~\cite{Cai:24}. Furthermore, the integration of, for example, time-bin-to-polarization encoders~\cite{Kupchak_2017} or, vice-versa, polarization-to-time bin ones~\cite{Scalcon2022} might allow for increased interoperability between different transmitters and receivers. From a theoretical perspective, the switching time should be optimized based on the current and foreseen free-space channel status, which is typically the most unstable, to allow for an equal distribution of the keys between the nodes and to improve the total network throughput.
Through this field trial, we demonstrated the feasibility of connecting remote areas and integrating different technologies in an already existing quantum network consisting of heterogeneous nodes, an operation which will be crucial for the deployment of the incoming quantum internet~\cite{kimble2008,doi:10.1126/science.aam9288}.

\subsection*{Authors' contributions}
The study conception and design were performed by F.~Picciariello, F.~Vedovato, M.~Avesani, G.~Foletto, G. Vallone, and P. Villoresi. Management and coordination responsibility for the research activity were performed by F.~Vedovato, M.~Avesani, G. Vallone and P. Villoresi. The experiment realization and the data collection were performed by F.~Picciariello, E.~Rossi, F.~Vedovato, M.~Avesani, and G.~Foletto. The data analysis was performed by I.~Karakosta-Amarantidou, F.~Picciariello, G.~Foletto, M.~Avesani, and F.~Vedovato. The software used was developed by F.~Picciariello, G.~Foletto, M.~Avesani, L.~Calderaro. The following authors contributed to the realization of the technical equipment used during the experiment: F.~Picciariello, I.~Karakosta-Amarantidou, G.~Foletto, M.~Avesani, E.~Rossi, L.~Calderaro, and F.~Vedovato. The first draft of the manuscript was written by I.~Karakosta-Amarantidou, F.~Picciariello, F.~Vedovato, G.~Foletto, and M.~Avesani. Acquisition of the financial support for the project leading to this publication was performed by P.~Villoresi. All authors read, contributed and approved the final manuscript.

\subsection*{Acknowledgements}
\label{sec:ack}
We would like to thank Flavio Seno, Marcello Lunardon and Ivan Rossatelli of the Department of Physics and Astronomy
for the collaboration and support at DFA, Gaetano Maron of LNL-INFN for the collaboration and support at LNL, as well as Francesca Bettini, Fabio Luise and Lorenzo Franceschin of the Department of Information Engineering for the
collaboration and support at DEI. We would like to thank also the projects Q-SecGround Space of the Italian Space
Agency, QUANCOM of European Union’s PON Research and Innovation programme, QUID of European Union’s Digital
programme, AppQInfo of European Union’s Horizon 2020 Research and Innovation programme, PON Ricerca e
Innovazione 2014-2020 of European Union for providing the preliminary research basis for this work.

\subsection*{Funding}
This project has received funding from the European Union’s Horizon Europe research and innovation programme under
the project “Quantum Secure Networks Partnership” (QSNP, grant agreement No 101114043).

{\it Disclaimer.} Funded by the European Union. Views and opinions expressed are however those of the author(s) only and do not necessarily reflect those of the European Union or European Commission-EU. Neither the European Union nor the granting authority can be held responsible for them.

\end{document}